\documentclass[]{article}

\usepackage{widetext}
\usepackage{graphicx}
\usepackage{caption}
\usepackage{subcaption}

\usepackage{fullpage}

\usepackage{verbatim}
\usepackage{amsmath}
\usepackage{amssymb}
\usepackage{color}
\usepackage{hyperref}
\usepackage{stackrel}
\usepackage{upgreek}
\usepackage{csquotes}

 \def\##1{{\bf #1}}
\renewcommand\vec{\mathbf}
\def\doubleunderline#1{\underline{\underline{#1}}}
\def\=#1{\underline{\underline{#1}}}

\def\magenta{\textcolor{black}}
 \def\eps{\varepsilon}

 \def\epso{\eps_{\scriptscriptstyle 0}}

\def\muo{\mu_{\scriptscriptstyle 0}}
\def\ko{k_{\scriptscriptstyle 0}}
\def\koc{{\ko}c}

\def\etao{\eta_{\scriptscriptstyle 0}}

\def\epsr{\eps_{r}}
\def\mur{\mu_{r}}

\def\.{\mbox{ \tiny{$^\bullet$} }}

\let\oldhat\hat
\renewcommand{\hat}[1]{\oldhat{\mathbf{#1}}}
\def\ux{\hat{\#x}}
\def\uy{\hat{\#y}}
\def\uz{\hat{\#z}}

\def\uphi{\hat{ \boldsymbol\phi}}
\def\utheta{\hat{ \boldsymbol\theta}}
\newenvironment{rcases}
  {\left.\begin{aligned}}
  {\end{aligned}\right\rbrace}
  
  \def\les{\left[}
\def\ris{\right]}
\def\lec{\left\{}
\def\ric{\right\}}
  
\def\GerM{\mbox{\boldmath${\mathfrak M}$}}
\def\GerN{\mbox{\boldmath${\mathfrak N}$}}

\def\newJ{{\mathfrak J}}
\def\newK{{\mathfrak K}}
\def\calP{{\cal P}}
\def\calQ{{\cal Q}}
\def\calR{{\cal R}}
\def\calU{{\cal U}}
\def\calV{{\cal V}}  

\def\st{\sin\theta}
\def\ct{\cos\theta}
\def\sp{\sin\phi}
\def\cp{\cos\phi}

\begin{document}

 \begin{center}
{\sf Planewave scattering by an ellipsoid composed of an orthorhombic
dielectric--magnetic material}\\
\vskip 0.5 cm

 {Hamad M. Alkhoori}$^{1}$
 {Akhlesh Lakhtakia}$^2$
{James K. Breakall},$^1$ and
{Craig F. Bohren}$^3$
\vskip 0.5 cm
 $^1${Department of Electrical Engineering, The Pennsylvania State University, University Park, Pennsylvania 16802, USA}\\
 $^2${Department of Engineering Science and Mechanics, The Pennsylvania State University, University Park, Pennsylvania 16802, USA}\\
$^3${Department of Meteorology, The Pennsylvania State University, University Park, Pennsylvania 16802, USA} 
 
\end{center}

\begin{abstract}
The extended boundary condition method (EBCM) can be used to study planewave scattering by an ellipsoid
composed of an orthorhombic dielectric-magnetic material whose relative permittivity dyadic is a scalar multiple of its relative permeability dyadic.  The  scattered and internal field phasors can be expanded in terms of appropriate vector spherical wavefunctions with  unknown expansion coefficients, whereas the  incident field phasors can be similarly expanded but with  known expansion coefficients. The scattered-field coefficients are related to the \magenta{incident-field} coefficients
through a \magenta{matrix. The} scattering, absorption, and extinction efficiencies \magenta{were} calculated thereby in relation to the propagation direction and the polarization state of the incident plane wave,  the constitutive-anisotropy parameters, and
the nonsphericity parameters of the \magenta{ellipsoid, when the} eigenvectors of the real permittivity dyadic are aligned along the three semi-axes of the ellipsoid. 
As the electrical size of the ellipsoid increases, multiple lobes appear in the scattering pattern.
The total scattering efficiency can be smaller than the absorption efficiency  for some configurations of the incident plane wave but not necessarily for others. The nonsphericity of the object has a stronger influence on the total scattering efficiency  than on the absorption efficiency. The forward-scattering efficiency increases monotonically with the electrical size for all configurations of the incident plane wave, and so does the backscattering efficiency for some configurations. For other configurations, the backscattering efficiency has an undulating behavior with increase in electrical size, and is  highly affected by the shape and the constitutive anisotropy of the ellipsoid.  Even though the ellipsoid is not necessarily a body of revolution, it is anisotropic, and it is not impedance matched to free space, the backscattering efficiency can be minuscule but the forward-scattering efficiency is not. This feature
can be useful for    harvesting electromagnetic energy.  
\end{abstract}

\section{Introduction}

The scattering of a time-harmonic electromagnetic field by a nonspherical object composed of a complex material is a topic of interest to contemporary researchers. Most natural objects are not spherical \cite{Thompson,Stanley} and many natural materials are not isotropic \cite{Klein,Gersten}. 

Many real problems require  analysis of electromagnetic fields in  anisotropic and bianisotropic materials \cite{EAB}. For example, many particles in planetary and interstellar dusts  are crystalline \cite{Molster,inter,Sokolik}. Thus, understanding the scattering characteristics of nonspherical crystalline objects may be  useful in inverse-scattering astrophysical and aerosol problems where information on the scattering object has to be determined from scattering data collected by a receiving antenna. Also, studies of the interplay of shape (i.e., nonsphericity) and material anisotropy/bianisotropy   can be useful in designing objects with desired scattering or absorption characteristics. One possible application is in the design of stealth sensors: whereas some absorption  must occur in any sensor, weak
 scattering is required for stealthy operation  
\cite{sensor}. Furthermore, nonspherical sensors can be more convenient for mounting on nonplanar surfaces.
Finally, the fabrication of new materials endowed with characteristics that are  unknown in nature has received considerable attention in the last few years. Examples are metamaterials \cite{Dudley, Tong} which are often fabricated by dispersing electrically small inclusions \cite{Hulst} in a host material. The scale of nonhomogeneity is controlled by properly adjusting the spacing between neighboring inclusions. In a particular spectral regime, the metamaterial can be considered to be an anisotropic or bianisotropic continuum \cite{Kriegler,MAEH}. 

Scattering by homogeneous 3D objects of a finite surface area  has long been of interest to the electromagnetics community \cite{Hulst,Bohren,Bowman,Hsiao,YlaOijala,Kunz,Monk}. The scattered fields may be analytically obtained by (i) expanding the incident, scattered, and internal fields \magenta{in terms of}
 suitable vector wavefunctions and (ii) imposing the appropriate boundary conditions at the surface of the scattering object, provided that one of the three coordinates of a coordinate system is constant on the surface and  the method of separation of variables 
can be used in that coordinate system to solve the frequency-domain Maxwell equations \cite{Bowman,Bohren, Beltrami}.
Due to these requirements, only boundary-value problems of
scattering by arbitrarily sized spheres and spheroids made of  isotropic  materials have been
solved in closed form \cite{Bohren,Bowman,Beltrami,Asano, Li}. Numerical techniques
are used for nonspherical objects \cite{Hsiao,YlaOijala,Kunz,Monk}.

An exception is  the extended boundary condition method (EBCM), also
called the   null-field method and the T-matrix method. This semi-analytical semi-numerical method was originally developed for scattering by an infinite-conductivity object by Waterman \cite{Waterman}, and was subsequently  extended to encompass objects made of  biisotropic materials  \cite{Lakh-IskBook}. This method requires   knowledge of (i) the bilinear expansions of the dyadic Green functions for the surrounding medium and (ii) closed-form vector wavefunctions to completely express the fields induced inside the object. 

The first requirement  was fulfilled decades ago for free space \cite{Tai}. The second requirement was fulfilled recently for  orthorhombic dielectric-magnetic materials obeying the frequency-domain constitutive relations \cite{Lakhtakiaoriginal}
 \begin{equation}
\begin{rcases}
\begin{aligned}
\vec{D}(\vec{r} )&= \epso \eps_r  \doubleunderline{A}  \. \doubleunderline{A}\. \vec{E}(\vec{r}) \\
\vec{B}(\vec{r} ) &= \muo \mur \doubleunderline{A}  \.\doubleunderline{A}\. \vec{H} (\vec{r})
\end{aligned}
\end{rcases},
\label{conrels}
\end{equation}   
 where $\epso$ and $\muo$ are the permittivity and permeability of free space, respectively; the
 diagonal dyadic
 \begin{equation}
 \doubleunderline{A}= \alpha_x^{-1} \, \ux\ux+ \alpha_y^{-1}\, \uy\uy+ \uz\uz\,;
 \label{A-def}
\end{equation}
$\eps_r$ and $\mur$ are complex functions of the angular frequency $\omega$; and 
 the constitutive-anisotropy
parameters $\alpha_x$ and $\alpha_y$ are real positive functions of $\omega$.  Thus, 
the relative permittivity dyadic 
\begin{equation}
\doubleunderline{\eps}_r=\epsr  \doubleunderline{A}  \. \doubleunderline{A}
\end{equation}
of this material is a scalar multiple of its relative permeability dyadic 
\begin{equation}
\doubleunderline{\mu}_r=\mur  \doubleunderline{A}  \. \doubleunderline{A}.
\end{equation}
The EBCM was used to investigate the planewave scattering characteristics of a sphere composed of this material \cite{Jafri,Jafri-err}. However,  scattering by a nonspherical object made of the same material has not been addressed yet.

Our aim for this paper was to examine the scattering of a plane wave by an ellipsoid composed of the material described by Eqs.~(\ref{conrels}) and (\ref{A-def}).  In a Cartesian coordinate system with its origin at the centroid of the ellipsoid,
the surface $S$ of the ellipsoid is delineated by the position vector
\begin{eqnarray}
&&
\nonumber
\#r_s(\theta,\phi)=c\=U\. \left[\left(\ux\cos\phi+\uy\sin\phi\right)\sin\theta +\uz\cos\theta\right]\,,
\\[5pt]
&&\quad
\quad \theta\in[0,\pi]\,,\quad\phi\in[0,2\pi)\,,
\end{eqnarray}
where
\begin{equation}
\=U=\left(a\ux\ux+b\uy\uy\right)/c+\uz\uz\,
\end{equation}
may be called the \textit{shape dyadic}.
 Thus, the ellipsoid   has linear
dimensions $2a, 2b$, and $2c$ along the $x$, $y$, and $z$ axes, respectively, and reduces to \magenta{a spheroid} if any two of the dimensions are equal or \magenta{a sphere} if all three are equal. The shape of the
ellipsoid is adequately described by the ratios $a/c$ and $b/c$ in $\=U$. 

The eigenvectors of $\=A$ and $\=U$ are identical. However, each of the two has at least two distinct  eigenvalues.
In order to study the interplay of shape  and constitutive anisotropy, we computed the differential scattering, total scattering, absorption, backscattering, and forward scattering cross sections \cite{Bohren,Jafri}.
The plan of the paper is as follows. In Section \ref{s1}, we present the EBCM equations for the chosen scattering problem, which is the scattering of a plane wave by an ellipsoid composed of an orthorhombic dielectric-magnetic material. In Section \ref{s2}, we present computed values of the various cross sections (after normalization by a
fixed area) in relation to the direction of propagation and the polarization state of the incident plane wave, the shape of the ellipsoid, and the  anisotropy of the ellipsoid material. Our conclusions are summarized in Section \ref{s3}. 
An $\exp(-i \omega t)$ dependence on time $t$ is implicit  throughout the analysis with $i=\sqrt{-1}$. Vectors are  in boldface, unit vectors are decorated by caret,  dyadics are  double underlined, and column vectors as well as matrices are enclosed in square brackets.

\section{Theory}\label{s1}
\subsection{Incident plane wave}
Let  the region occupied by a homogeneous ellipsoid be denoted by $V$; accordingly, $\#r(\theta,\phi)\in{V}\Rightarrow 
\vert\#r(\theta,\phi)\vert\le\vert\#r_S(\theta,\phi)\vert$. The region outside $V$ is vacuous. A plane wave is incident on the ellipsoid.
The electric  and magnetic field phasors of the incident plane wave are given as
\begin{equation}
\label{Einc}
\vec{E}_{inc}(\vec{r})= \hat{e}_{inc} \exp(i\vec{k}_{inc} \cdot \vec{r})
\end{equation}
and 
\begin{equation}
\vec{H}_{inc}(\vec{r})=\frac{\vec{k}_{inc}\times\hat{e}_{inc}}{\omega\muo}\, \exp\left(i\vec{k}_{inc}\.\#r\right)\,,
\end{equation}
respectively. Here the wave vector
\begin{equation}
\vec{k}_{inc}= \ko(\hat{x} \sin \theta_{inc} \cos \phi_{inc} + \hat{y} \sin \theta_{inc} \sin \phi_{inc}+\hat{z} \cos \theta_{inc})
\end{equation}
involves the angles $\theta_{inc}\in[0,\pi]$ and $\phi_{inc}\in[0,2\pi)$ defining the incidence direction, the
unit vector $\hat{e}_{inc}$ defines the polarization state, and $\ko=\omega\sqrt{\epso \muo}$ is the   free-space wavenumber. We also define the unit vectors $\hat{k}_{inc}=\vec{k}_{inc}/ \ko$ and
$\hat{h}_{inc}=\hat{k}_{inc}\times\hat{e}_{inc}$ for later convenience.

The incident electric and magnetic field phasors  may be expressed as
\begin{equation}
 \label{1}
\begin{aligned}
\vec{E}_{inc}(\vec{r}) =&  \lim_{N\to\infty} \sum_{s \in \{e,o\}} \sum_{n=1}^N \sum_{m=0}^n  \big\{
D_{mn}\big[A_{smn}^{(1)} \vec{M}_{smn}^{(1)}(\ko \vec{r}) \\
& + B_{smn}^{(1)} \vec{N}_{smn}^{(1)}(\ko \vec{r}) \big]\big\}\,
\end{aligned}
\end{equation}
and 
\begin{equation}
 \label{1-H}
\begin{aligned}
\vec{H}_{inc}(\vec{r}) =&-\frac{i}{\etao}  \lim_{N\to\infty} \sum_{s \in \{e,o\}} \sum_{n=1}^N \sum_{m=0}^n 
 \big\{ D_{mn}\big[ A_{smn}^{(1)} \vec{N}_{smn}^{(1)}(\ko \vec{r}) \\
& + B_{smn}^{(1)} \vec{M}_{smn}^{(1)}(\ko \vec{r}) \big]\big\}\,,
\end{aligned}
\end{equation}
respectively,
where $\etao= \sqrt{\muo/\epso}$ is the intrinsic impedance of free space.
The normalization factor
\begin{equation}
D_{mn} = (2- \delta_{m0}) \frac{(2n+1)(n-m)!}{4n(n+1)(n+m)!}
\end{equation}
involves the Kronecker delta $\delta_{m m^\prime}$.

The expansion coefficients  are given by \cite{Jafri, Morse}
\begin{equation}
\begin{rcases}
\begin{aligned}[b]
A_{smn}^{(1)} = 4 i^n \sqrt{n(n+1)}\, \hat{e}_{inc} \. \vec{C}_{smn}(\theta_{inc},\phi_{inc})  \\
B_{smn}^{(1)} = 4 i^{n-1} \sqrt{n(n+1)}\, \hat{e}_{inc} \. \vec{B}_{smn}(\theta_{inc},\phi_{inc}) 
\end{aligned}
\end{rcases},
\end{equation}
where the vector spherical harmonics
\begin{equation}
\vec{B}_{smn}(\theta,\phi)= \hat{r} \times \vec{C}_{smn}(\theta,\phi)
\end{equation}
and
\begin{eqnarray}
\nonumber
&&
 \vec{C}_{smn}(\theta,\phi)=
  \frac{1}{\sqrt{n(n+1)}} \left[ \mp m \frac{P_n^m(\cos \theta)}{\sin \theta}  \Bigg\{ \begin{matrix}
\sin (m \phi)\\
\cos (m \phi)
\end{matrix} \Bigg\} \utheta \right.
\\[8pt]
&& \quad\left.-\frac{d P_n^m(\cos \theta)}{d \theta} \Bigg\{ \begin{matrix}
\cos (m\phi)\\
\sin (m\phi)
\end{matrix} \Bigg\}  \uphi  \right]  \,,\quad
s=\left\{\begin{array}{l} e \\ o\end{array}\right.
\end{eqnarray}
involve the associated Legendre function $P_n^m(\cos \theta)$ of order $n$ and degree $m$,
and the index $s$ stands for either  even (e) or odd (o) parity. 

The vector spherical wavefunctions of the first kind,
$\vec{M}_{smn}^{(1)}(\ko \vec{r})$ and $\vec{N}_{smn}^{(1)}(\ko \vec{r})$, are  available in standard texts \cite{Morse,Stratton}, 
the  index $n$ denoting the order of the spherical Bessel function $j_n(\ko{r})$ appearing in those wavefunctions.
The index $n$ is restricted to $[1,N]$ where $N$ is sufficiently large and the limit on the right sides of Eqs.~(\ref{1}) and (\ref{1-H})  are not used.  

\subsection{Scattered field}

 The scattered electric and magnetic field phasors take the form
 \begin{equation}
 \label{2}
\begin{aligned}
\vec{E}_{sca}(\vec{r}) =&  \lim_{N\to\infty} \sum_{s \in \{e,o\}} \sum_{n=1}^N \sum_{m=0}^n
\big\{  D_{mn}\big[ A_{smn}^{(3)} \vec{M}_{smn}^{({3})}(\ko \vec{r}) \\
& + B_{smn}^{(3)} \vec{N}_{smn}^{({3})}(\ko \vec{r}) \big]\big\}\,
\end{aligned}
\end{equation}
and 
 \begin{equation}
 \label{2-H}
\begin{aligned}
\vec{H}_{sca}(\vec{r}) =&-\frac{i}{\etao}  \lim_{N\to\infty} \sum_{s \in \{e,o\}} \sum_{n=1}^N \sum_{m=0}^n
\big\{  D_{mn}\big[ A_{smn}^{(3)} \vec{N}_{smn}^{({3})}(\ko \vec{r}) \\
& + B_{smn}^{(3)} \vec{M}_{smn}^{({3})}(\ko \vec{r}) \big]\big\}\,,
\end{aligned}
\end{equation}
respectively.
The vector spherical wavefunctions of the third kind \cite{Morse,Stratton}, $\vec{M}_{smn}^{(3)}(\ko \vec{r}) $ and $\vec{N}_{smn}^{(3)}(\ko \vec{r}) $, involve the spherical Hankel function $h_n^{(1)}(\ko{r})$ instead of $j_n(\ko{r})$. The unknown expansion coefficients $A_{smn}^{(3)}$ and $B_{smn}^{(3)}$ have to be determined. 
The scattered field phasors thus contain \magenta{magnetic-multipole} terms quantified by the coefficients
$A_{smn}^{(3)}$ and \magenta{electric-multipole} terms quantified by the coefficients
$B_{smn}^{(3)}$ \cite{Jackson}.

By making use of the Ewald--Oseen extinction theorem and exploiting the orthogonality properties of
the vector spherical wavefunctions \cite{Lakhtakiaoriginal}, the \magenta{incident-field} coefficients
and the scattered field coefficients can be related to the tangential components of the electric and magnetic field phasors on $S$; accordingly,
\begin{equation} 
\label{3-A}
\begin{aligned}
A_{smn}^{(j)} =& \mp\frac{i \ko^2}{\pi} \iint_S \big\{ \left[ \hat{n}(\vec{r}_s) \times \vec{E}_{int}(\vec{r}_s) \right] \.\vec{N}_{smn}^{(\ell)}(\ko \vec{r}_s)  \\
& + i \etao \left[ \hat{n}(\vec{r}_s) \times \vec{H}_{int}(\vec{r}_s) \right]\. \vec{M}_{smn}^{(\ell)}(\ko \vec{r}_s) \big\} d^2 \vec{r}_s
\end{aligned}
\end{equation}
and
\begin{equation} 
\label{3-B}
\begin{aligned}
B_{smn}^{(j)} =& \mp \frac{i \ko^2}{\pi} \iint_S  \big\{ \left[ \hat{n}(\vec{r}_s) \times \vec{E}_{int}(\vec{r}_s) \right] \. \vec{M}_{smn}^{(\ell)}(\ko \vec{r}_s) \\
&+ i \etao \left[\hat{n}(\vec{r}_s) \times \vec{H}_{int}(\vec{r}_s) \right]\. \vec{N}_{smn}^{(\ell)}(\ko \vec{r}_s) \big\}d^2 \vec{r}_s \,.
\end{aligned}
\end{equation}
Here, $\hat{n}(\vec{r}_s)=\nabla{r_s(\theta,\phi)}/\vert\nabla{r_s(\theta,\phi)}\vert$
is the unit outward normal to $S$ at $\vec{r}_s \in S$, $j \in [1,3]$, and $\ell=j+2( \text{mod} 4) \in [3,1]$. The upper signs are used on the left sides
of Eqs.~(\ref{3-A}) and (\ref{3-B}) when $j=1$, the lower signs when $j=3$.\\

\subsection{Internal field}
The electric and magnetic field phasors excited inside the scattering object are represented by  \cite{Lakhtakiaoriginal}
\begin{eqnarray}
\nonumber
&&\vec{E}_{int}(\#r)=\lim_{N\to\infty}   \sum_{s\in\left\{e,o\right\}}\sum^{N}_{n=1}\sum^n_{m=0}
\left[
b_{smn}  \,\GerM_{smn}(\#r)\right.
\\[5pt]
&&\qquad+\left.
c_{smn}  \,\GerN_{smn}(\#r)\right]\,
\label{Eint}
\end{eqnarray}
and
\begin{eqnarray}
\nonumber
&&\vec{H}_{int}(\#r)=
-\frac{i}{\etao}\sqrt{\frac{\epsr}{\mur}}\lim_{N\to\infty}
  \sum_{s\in\left\{e,o\right\}}\sum^{N}_{n=1}\sum^n_{m=0}
\left[
b_{smn}  \,\GerN_{smn}(\#r)\right.
\\[5pt]
&&\qquad
+\left.
c_{smn}  \,\GerM_{smn}(\#r)\right]\,, 
\label{Hint}
\end{eqnarray}
where the expansion coefficients $b_{smn}$ and $c_{smn}$ are not known.
The functions $\GerM_{smn}(\vec{r})$ and $\GerN_{smn}(\vec{r})$ are defined as  
\newpage
\begin{widetext}
\begin{eqnarray}
\nonumber
&&\GerM_{smn}(\#r)=
\frac{\newJ_n(k\#r)}{f_1(\phi)}\,\doubleunderline{A}^{-1}\.
\\
&&\nonumber\quad
\lec
\hat{\#r} \left[
\frac{f_4(\phi)-f_1^2(\phi)}{f_2(\theta,\phi)}\st\ct  \,\calQ_{smn}(\theta,\phi)
-(\alpha_x-\alpha_y)\st\sp\cp \,\calR_{smn}(\theta,\phi)\right]
\right.
\\
&&\nonumber\qquad
+
\utheta \left[
\frac{f_4(\phi)\cos^2\theta+f_1^2(\phi)\sin^2\theta}{f_2(\theta,\phi)} \,\calQ_{smn}(\theta,\phi)
-(\alpha_x-\alpha_y)\ct\sp\cp\,\calR_{smn}(\theta,\phi)\right]
\\
&&\qquad
+
\left.
\uphi \left[
-\,\frac{\alpha_x-\alpha_y}{f_2(\theta,\phi)}\ct\sp\cp \,\calQ_{smn}(\theta,\phi)
- f_4(\phi)  \,\calR_{smn}(\theta,\phi)\right]
\ric
\,
\end{eqnarray}
and
\begin{eqnarray}
\nonumber
&&\GerN_{smn}(\#r)=
\doubleunderline{A}^{-1}
\.\Bigg(
\hat{\#r}\lec \frac{\newJ_n(k\#r)}{kr }\,
\les\frac{\cos^2\theta+f_4(\phi)\sin^2\theta}{f_2^2(\theta,\phi)}\ris \,
\calP_{smn}(\theta,\phi)\,
\right.\\ &&\nonumber\qquad\quad\left.
+\frac{\newK_n(k\#r)}{f_1(\phi)}
\left[
\frac{f_4(\phi)-f_1^2(\phi)}{f_2(\theta,\phi)}\st\ct \,\calR_{smn}(\theta,\phi)  +
(\alpha_x-\alpha_y)\st\sp\cp\,\calQ_{smn}(\theta,\phi)\right]\ric
\\
&&
\nonumber\qquad
+\utheta\lec\frac{\newJ_n(k\#r)}{kr}\,
\les\frac{f_4(\phi)-1}{ f_2^2(\theta,\phi)}\st\ct\ris\,
\calP_{smn}(\theta,\phi)\,
\right.\\ &&\nonumber\qquad\quad\left.
+\frac{\newK_n(k\#r)}{f_1(\phi)}
\left[
\frac{f_4(\phi)\cos^2\theta+f_1^2(\phi)\sin^2\theta}{f_2(\theta,\phi)}\,
\calR_{smn}(\theta,\phi)  +
(\alpha_x-\alpha_y)\ct\sp\cp\,\calQ_{smn}(\theta,\phi)\right]\ric
\\
&&\nonumber\qquad
+\uphi\lec
-\,\frac{\newJ_n(k\#r)}{kr}\,
\les\frac{\alpha_x-\alpha_y}{f_2^2(\theta,\phi)}\st\sp\cp\ris\,
\calP_{smn}(\theta,\phi)\,
\right.\\ && \qquad\quad\left.
+\frac{\newK_n(k\#r)}{f_1(\phi)}
\left[
-\,\frac{\alpha_x-\alpha_y}{f_2(\theta,\phi)}\ct\sp\cp\,\calR_{smn}(\theta,\phi)
+f_4(\phi)\,\calQ_{smn}(\theta,\phi)\right]\ric\Bigg)\,,
\end{eqnarray}
\end{widetext}
where
\begin{eqnarray}
&&
k=\ko\frac{\sqrt{\epsr}\sqrt{\mur}}{\alpha_x\alpha_y}\,,
\\
&&
\newJ_n(k\#r)=j_n\left[krf_2(\theta,\phi)\right]\,,
\\
&&
\newK_n(k\#r)=\frac{n+1}{krf_2(\theta,\phi)}\, \newJ_n(k\#r) -\newJ_{n+1}(k\#r)\,,
\\
&&
\calP_{smn}(\theta,\phi)=n(n+1)P_n^m\left[\frac{\cos\theta}{f_2(\theta,\phi)}\right]\calV_{sm}(\phi)\,,
\\
&&
\calQ_{smn}(\theta,\phi)=mP_n^m\left[\frac{\cos\theta}{f_2(\theta,\phi)}\right]\frac{f_2(\theta,\phi)}{f_1(\phi)\sin\theta}\,
\calU_{sm}(\phi)\,,
\\
&&\calR_{smn}(\theta,\phi)=\frac{1}{f_1(\phi)\sin\theta}
\nonumber
\\
\nonumber
&&\quad\magenta{\times}
\left\{
(n-m+1) {f_2(\theta,\phi)} P_{n+1}^m\left[\frac{\cos\theta}{f_2(\theta,\phi)}\right]
\right.
\\
&&
\qquad
\left.
-
(n+1) {\cos\theta} {P_n^m}\left[\frac{\cos\theta}{f_2(\theta,\phi)}\right]\right\}\calV_{sm}(\phi)\,,
\\
&&
\calU_{sm}(\phi)=\lec\begin{array}{c}-\sin \left[mf_3(\phi)\right]\\\cos\left[mf_3(\phi)\right]\end{array}\ric
\,,
\quad s=\left\{\begin{array}{c}e\\{o}\end{array}\right.\,,
\\
&&
\calV_{sm}(\phi)=\lec\begin{array}{c}\cos \left[mf_3(\phi)\right]\\\sin\left[mf_3(\phi)\right]\end{array}\ric
\,,
\quad s=\left\{\begin{array}{c}e\\{o}\end{array}\right.\,,
\\
&&
f_1(\phi) =+\left(\alpha_x^2\cos^2\phi+\alpha_y^2\sin^2\phi\right)^{1/2}\,,
\end{eqnarray}
\begin{eqnarray}
\nonumber
&&f_2(\theta,\phi) =+\left[f_1^2(\phi)\sin^2\theta+\cos^2\theta\right]^{1/2}\,,
\\
&&
f_3(\phi)=\tan^{-1}\left(\frac{\alpha_y}{\alpha_x}\tan\phi\right)\,,
\\
&&
f_4(\phi)=\alpha_x \cos^2\phi+\alpha_y \sin^2\phi\,.
\end{eqnarray}
The angle $f_3(\phi)$ must lie in the same quadrant as its argument.

\subsection{T matrix}

After substituting Eqs.~(\ref{Eint}) and (\ref{Hint}) in Eq.~(\ref{3-A}) and (\ref{3-B}) with
$n\leq {N}$, a set of algebraic equations emerges to relate
the scattered-field coefficients to the incident-field coefficients. Symbolically, this relationship
is expressed in matrix form as \cite{Waterman,Lakh-IskBook}
\begin{equation}
\begin{pmatrix}
A_{smn}^{(3)} \\ ---- \\ B_{smn}^{(3)}
\end{pmatrix}
=[T] 
\begin{pmatrix}
A_{smn}^{(1)} \\ ---- \\ B_{smn}^{(1)}
\end{pmatrix},
\end{equation} 
where 
\begin{equation} 
\label{4a}
[T]= -[Y^{(3)}] [Y^{(1)}]^{-1}
\end{equation}
is the T matrix.

The matrix $[Y^{(j)}]$, $j\in[1,3]$, is symbolically written as
\begin{equation} 
\label{5}
[Y^{(j)}]= 
\begin{pmatrix}
I_{smn,s^\prime m^\prime n^\prime}^{(j)} && \big| && J_{smn,s^\prime m^\prime n^\prime}^{(j)} \\
---- && \big| && ---- \\
K_{smn,s^\prime m^\prime n^\prime}^{(j)}&& \big| && L_{smn,s^\prime m^\prime n^\prime}^{(j)}
\end{pmatrix}.
\end{equation}
\newpage
\begin{widetext}
The matrix elements in Eq. (\ref{5}) are double integrals given by 
\begin{equation} 
\label{6-I}
\begin{aligned}[b]
I_{smn,s^\prime m^\prime n^\prime}^{(j)} =& -\frac{i \ko^2}{\pi} 
\int \limits_{\phi=0}^{2 \pi}   \int \limits_{\theta=0}^\pi   
r_s^2(\theta, \phi)  \vert\nabla r_s(\theta, \phi) \vert \sin \theta
\Bigg[
\lec\vec{N}_{smn}^{(\ell)}(\ko \vec{r}_s) \. 
\les \hat{n}(\vec{r}_s)  \times  \GerM_{s^\prime m^\prime n^\prime}(\vec{r}_s)  \ris\ric
\\
+ &\sqrt{\frac{\epsr}{\mur}}
\lec\vec{M}_{smn}^{(\ell)}(\ko \vec{r}_s) \. 
\les \hat{n}(\vec{r}_s)  \times  \GerN_{s^\prime m^\prime n^\prime}(\vec{r}_s) \ris\ric 
\Bigg]d\theta\,d\phi\,,
\end{aligned}
\end{equation}
\begin{equation} 
\label{6-J}
\begin{aligned}[b]
J_{smn,s^\prime m^\prime n^\prime}^{(j)} =& -\frac{i \ko^2}{\pi} 
\int \limits_{\phi=0}^{2 \pi}   \int \limits_{\theta=0}^\pi   
r_s^2(\theta, \phi)  \vert\nabla r_s(\theta, \phi) \vert \sin \theta
\Bigg[
\lec\vec{N}_{smn}^{(\ell)}(\ko \vec{r}_s) \. 
\les \hat{n}(\vec{r}_s)  \times  \GerN_{s^\prime m^\prime n^\prime}(\vec{r}_s)  \ris\ric
\\
+ &\sqrt{\frac{\epsr}{\mur}}
\lec\vec{M}_{smn}^{(\ell)}(\ko \vec{r}_s) \. 
\les \hat{n}(\vec{r}_s)  \times  \GerM_{s^\prime m^\prime n^\prime}(\vec{r}_s) \ris\ric 
\Bigg]d\theta\,d\phi\,,
\end{aligned}
\end{equation}
\begin{equation} 
\label{6-K}
\begin{aligned}[b]
K_{smn,s^\prime m^\prime n^\prime}^{(j)} =& -\frac{i \ko^2}{\pi} 
\int \limits_{\phi=0}^{2 \pi}   \int \limits_{\theta=0}^\pi   
r_s^2(\theta, \phi)  \vert\nabla r_s(\theta, \phi) \vert \sin \theta
\Bigg[
\lec\vec{M}_{smn}^{(\ell)}(\ko \vec{r}_s) \. 
\les \hat{n}(\vec{r}_s)  \times  \GerM_{s^\prime m^\prime n^\prime}(\vec{r}_s)  \ris\ric
\\
+ &\sqrt{\frac{\epsr}{\mur}}
\lec\vec{N}_{smn}^{(\ell)}(\ko \vec{r}_s) \. 
\les \hat{n}(\vec{r}_s)  \times  \GerN_{s^\prime m^\prime n^\prime}(\vec{r}_s) \ris\ric 
\Bigg]d\theta\,d\phi\,,
\end{aligned}
\end{equation}
and
\begin{equation} 
\label{6-L}
\begin{aligned}[b]
L_{smn,s^\prime m^\prime n^\prime}^{(j)} =& -\frac{i \ko^2}{\pi} 
\int \limits_{\phi=0}^{2 \pi}   \int \limits_{\theta=0}^\pi   
r_s^2(\theta, \phi)  \vert\nabla r_s(\theta, \phi) \vert \sin \theta
\Bigg[
\lec\vec{M}_{smn}^{(\ell)}(\ko \vec{r}_s) \. 
\les \hat{n}(\vec{r}_s)  \times  \GerN_{s^\prime m^\prime n^\prime}(\vec{r}_s)  \ris\ric
\\
+ &\sqrt{\frac{\epsr}{\mur}}
\lec\vec{N}_{smn}^{(\ell)}(\ko \vec{r}_s) \. 
\les \hat{n}(\vec{r}_s)  \times  \GerM_{s^\prime m^\prime n^\prime}(\vec{r}_s) \ris\ric 
\Bigg]d\theta\,d\phi\,.
\end{aligned}
\end{equation}
\end{widetext}

The integrals in Eqs. (\ref{6-I})--(\ref{6-L}) can be obtained analytically only for an isotropic dielectric-magnetic sphere (i.e., $\alpha_x=\alpha_y=1$ and $a=b=c$) because then $\GerM_{smn}(\vec{r})$ reduces to $\vec{M}^{(1)}_{smn}(k\vec{r})$
and $\GerN_{smn}(\vec{r})$  to $\vec{N}^{(1)}_{smn}(k\vec{r})$. We used the Gauss--Legendre quadrature scheme
\cite{Jaluria} to evaluate these integrals. By testing against known integrals \cite{GS},  the numbers of nodes for integration over $\theta$ and $\phi$ were chosen to deliver the integrals correct to $\pm0.1\%$ relative error.

\subsection{Scattering, absorption, and extinction efficiencies}\label{sae}
Sufficiently far away from the object, the scattered electric field phasor
can be approximated as \cite{Bowman,Bohren}
\begin{equation}
\vec{E}_{sca}(r,\theta,\phi) \approx \vec{F}_{sca}( \theta,\phi) \frac{\exp(i\ko r)}{r},
\end{equation}
where \cite{Jafri}
\begin{equation}
\begin{aligned}[b]
\vec{F}_{sca}(\theta,\phi)=& \frac{1}{\ko}\lim_{N\to\infty}
 \sum_{s \in \{e,o\}} \sum_{n=1}^{N} \sum_{m=0}^n \bigg\{(-i)^n D_{mn} \sqrt{n(n+1)} \\
& \big[ -i A_{smn}^{(3)} \vec{C}_{smn}(\theta,\phi) + B_{smn}^{(3)} \vec{B}_{smn}(\theta, \phi) \big] \bigg\}.
\end{aligned}
\end{equation}
This quantity is useful in defining the differential scattering cross section
\begin{equation}
\label{sigmaD-def}
\sigma_D(\theta,\phi)= 4 \pi  \vert\vec{F}_{sca}(\theta,\phi)\vert^2 \,,
\end{equation}
whence the forward scattering cross section
\begin{equation}
\sigma_f= \sigma_D(\theta_{inc},\phi_{inc})\,,
\end{equation}
 the backscattering cross section
\begin{equation}
\sigma_b= \sigma_D(\pi+ \theta_{inc}, \pi + \phi_{inc})\,,
\end{equation}
and the extinction cross section 
\begin{equation}
 \sigma_{ext}=  \frac{4 \pi}{\ko} {\rm Im} \left[{\vec{F}_{sca}(\theta_{inc},\phi_{inc})\.  \hat{e}_{inc}^\ast}\right]
\end{equation}
follow, the asterisk indicating the complex conjugate.

By integrating $\sigma_D(\theta,\phi)$ over the entire solid angle, the total scattering cross section is
obtained as
\begin{equation}
\sigma_{sca}=  \frac{\pi}{\ko^2} \sum_{s \in \{e,o\}} \sum_{n=1}^{N} \sum_{m=0}^n  D_{mn} \left[ \vert{A_{smn}^{(3)}}\vert^2+ \vert{B_{smn}^{(3)}}\vert^2 \right]\,.
\end{equation}
Finally, the absorption cross section can be calculated as \cite{Garner-PRA}
\begin{equation}
 \sigma_{abs}= \sigma_{ext}- \sigma_{sca}\,.
 \end{equation}
 
Every cross section  defined in this section was divided by  $\pi {c^2}$ to convert it into a
dimensionless quantity called efficiency: $Q=\sigma/\pi{c^2}$.

\section{Numerical Results and Discussion} \label{s2}
A Mathematica\texttrademark~program was written to compute the T matrix using the
\magenta{lower-upper decomposition}   method to  invert $[Y^{(1)}]$ \cite{math}.
The value of $N$
was incremented by unity until the backscattering efficiency
$Q_b$ converged within a tolerance of $\pm0.1\%$. Of all the efficiencies defined in Sec.~2.\ref{sae},
$Q_b$ took the longest to converge.
 The larger the deviation of $k/\ko$ from unity, the higher was the value
 of $N$ required to achieve convergence.  The highest value of $N$ is $9$ for all results reported here.

Two conventions exist to define associated Legendre functions. These can be denoted as $P_n^m(\cos\theta)$ and $(-1)^m P_n^m(\cos\theta)$.  The associated Legendre functions native to Mathematica\texttrademark~ have to be multiplied by $(-1)^m$ in order to obtain the ones provided by Morse and Feshbach \cite[][pp.~1920--1921]{Morse} and used by us.

 Validation of the program was accomplished by checking against results available for simpler problems. The first validation was performed against the Lorenz--Mie theory for isotropic dielectric-magnetic spheres \cite{Stratton}. Regardless of the 
 \magenta{incidence} direction, all efficiencies were the same as available in the literature \cite{Bohren}. For an anisotropic sphere made of a material described by Eqs.~(\ref{conrels}) and (\ref{A-def}), our program was completely in accord with published data \cite{Jafri,Jafri-err}. 
 
The fields scattered by an electrically small ellipsoid made of the
material described by Eqs.~(\ref{conrels}) and (\ref{A-def}) \magenta{were  correctly} delivered by our program  \cite{Lakh-BBPC}. Also, our program agreed with the analytical conclusion that  scattering by a sphere made of an orthorhombic dielectric/magnetic material is equivalent to scattering by an ellipsoid made of an isotropic dielectric
(resp. magnetic) material, both objects being electrically small, provided that certain conditions are met; see the Appendix.
 
Convergence issues required attention for isotropic dielectric-magnetic spheroids. 
For prolate spheroids of aspect ratio $2$ (i.e., $c/a=c/b=2$) and oblate spheroids of aspect ratio $0.5$
(i.e., $c/a=c/b=0.5$), both nonmagnetic (i.e., $\mu_r=1$) the results generated by our program agreed with those of  Asano 
and Yamamoto \cite{Asano}, regardless of the size parameter $\ko{c}$. However, for prolate spheroids of aspect ratio $5$  and oblate spheroids of aspect ratio $0.2$, the scattering patterns were not in acceptable agreement with those of
Asano 
and Yamamoto \cite{Asano} for $\ko{c}>4$. The disagreement is rooted in the implicit reliance of the EBCM on analytic continuation \cite{Barber,convergence,Werby,LVVmag}, which becomes unstable in practice \cite{Lewin,Lefschetz}. 
Whereas analytic continuation of the electric and magnetic field phasors
everywhere inside $V$ is guaranteed by virtue of the frequency-domain Maxwell equations, the analytic continuation of 
the right sides of Eqs. (\ref{Eint}) and (\ref{Hint}) for finite $N$ is not guaranteed. For highly aspherical objects, 
analytic continuation for finite $N$ amounts to the supergain problem and leads to the ill conditioning of  $[Y^{(1)}]$
as $\ko$ increases \cite{EBCM-review}.
 Thus, EBCM by itself is   appropriate only for nonspherical objects that do not deviate too much from  a sphere. 

Nevertheless, several modifications can be applied to overcome the convergence problem \cite{convergence,LVVmag,EBCM-review}. With one of these modifications---viz,  reinforced orthogonalization of  $[Y^{(1)}]$ \cite{high} for a nondissipative object---our program yielded results in total agreement with published ones for highly aspherical ellipsoids \cite{high}.

 Parenthetically, both the iterative EBCM \cite{convergence} and the invariant imbedding T-matrix method 
\cite{iitm1, iitm2} are
improvements over the EBCM by itself for handling more aspherical and electrically larger scatterers.
But, as their implementation is computationally intricate even for isotropic scatterers,  we chose the  EBCM in order to focus on the effects of
 material anisotropy  while keeping the solution procedure as simple as possible.

In the remainder of this section, we present illustrative numerical results on the scattering, absorption, and extinction efficiencies of biaxially dielectric-magnetic ellipsoids ($a\ne{b}\ne{c}$ and $\alpha_x\ne\alpha_y\ne1$) and 
uniaxially dielectric-magnetic spheroids ($a={b}\ne{c}$ and $\alpha_x=\alpha_y\ne1$) in relation to 
\begin{itemize}
\item  the  propagation direction  of the incident plane wave ($\hat{k}_{inc}$),
\item the polarization state of the incident plane wave ($\hat{e}_{inc}$),
\item  the constitutive-anisotropy parameters  $\alpha_x$ and $\alpha_y$, 
\item  the nonsphericity parameters $a/c$ and $b/c$, and
\item  the electrical size $\koc$ of the semi-major axis.
\end{itemize}

\subsection{Differential scattering efficiency}
Although our program can accommodate any  incident plane wave, we limit our results to $\hat{k}_{inc}=\hat{z}$ and $\hat{e}_{inc}\in\left\{\hat{x},\hat{y}\right\}$. With $\epsr=4$  and $\mur=1.1$ fixed, we also set
\begin{itemize}
\item[(i)] $a/c=1/2$, $b/c=2/3$, $\alpha_x=1.1$, and $\alpha_y=1.2$  for the  biaxially dielectric-magnetic ellipsoid,
and
\item[(ii)] $a/c=b/c=1/2$ and $\alpha_x= \alpha_y=1.1$  for the  uniaxially dielectric-magnetic  spheroid.
\end{itemize}
The differential scattering efficiency $Q_D(\theta,\phi)$ was examined as a function of $\theta$  for 
$\phi\in\left\{0^\circ,90^\circ\right\}$, there being a twofold symmetry in the $xy$ plane, i.e., $Q_D(\theta,\phi+\pi)=Q_D(\theta,\phi)$. 
Plots of $Q_D(\theta,\phi)$ vs. $\theta$ for fixed $\phi$ are often called \textit{scattering patterns}.

\begin{figure}[h]
 \centering 
     \begin{subfigure}[h]{0.4\textwidth}
\includegraphics[width=\linewidth]{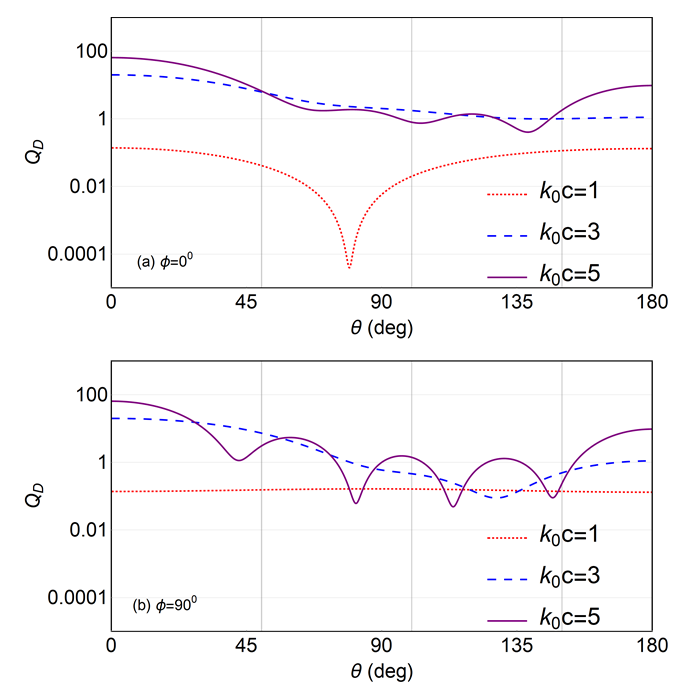}
 \end{subfigure}
\caption{$Q_D(\theta,\phi)$ vs. $\theta$ for  a biaxially dielectric-magnetic  ellipsoid when $\hat{k}_{inc}=\hat{z}$ and
$\hat{e}_{inc}=\hat{x}$; $\epsr=4$, $\mur=1.1$,
$a/c=1/2$, $b/c=2/3$, $\alpha_x=1.1$, and $\alpha_y=1.2$. The red dotted lines represent $\koc=1$, the blue dashed lines  $\koc=3$, and the purple solid lines  $\koc=5$.  (a) $\phi=0^\circ$, (b) $\phi=90^\circ$.}
\label{QD-biax-ell-Ex}
\end{figure}

\begin{figure}[h]
  \centering
     \begin{subfigure}[h]{0.4\textwidth}
\includegraphics[width=\linewidth]{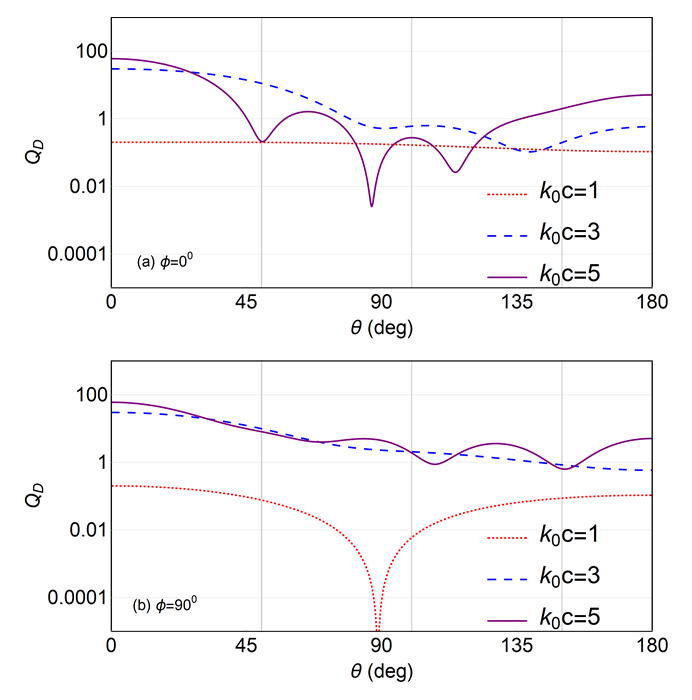}
 \end{subfigure}
\caption{Same as Fig. \ref{QD-biax-ell-Ex}, but when $\hat{e}_{inc}=\hat{y}$.
}
\label{QD-biax-ell-Ey}
\end{figure}

\begin{figure}[h]
  \centering
     \begin{subfigure}[h]{0.4\textwidth}
\includegraphics[width=\linewidth]{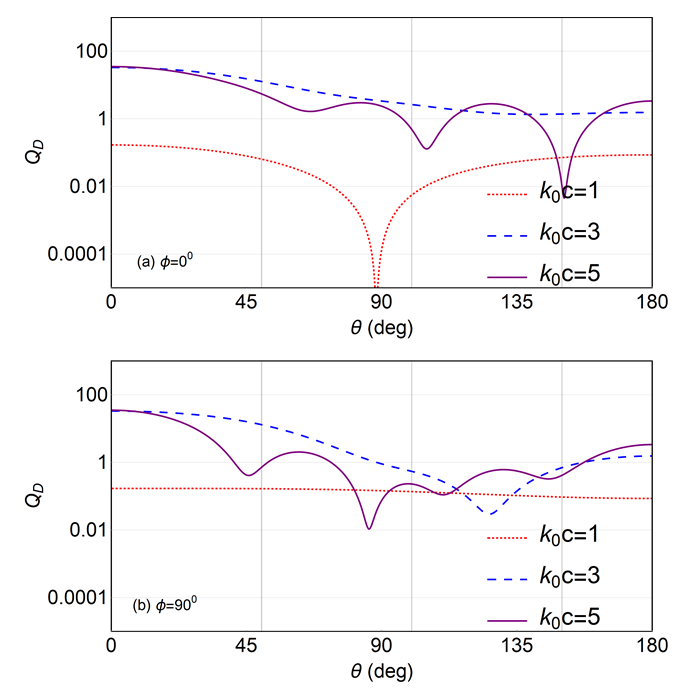}
 \end{subfigure}
\caption{$Q_D(\theta,\phi)$ vs. $\theta$ for a 
uniaxially dielectric-magnetic spheroid when $\hat{k}_{inc}=\hat{z}$; $\epsr=4$, $\mur=1.1$,
$a/c=b/c=1/2$, and $\alpha_x=\alpha_y=1.1$. The red lines represent $\koc=1$, the blue lines  $\koc=3$, and the purple lines  $\koc=5$. (a) Either $\lec\phi=0^\circ,\hat{e}_{inc}=\hat{x}\ric$ or $\lec\phi=90^\circ,\hat{e}_{inc}=\hat{y}\ric$.
(b) Either $\lec\phi=90^\circ,\hat{e}_{inc}=\hat{x}\ric$ or $\lec\phi=0^\circ,\hat{e}_{inc}=\hat{y}\ric$.
}
\label{QD-uniax-sph-Ex}
\end{figure}

Scattering patterns for $\phi\in\lec0^\circ,90^\circ\ric$ and $\koc\in\lec1,3,5\ric$ of the chosen  biaxially dielectric-magnetic  ellipsoid are depicted in Figs. \ref{QD-biax-ell-Ex} and \ref{QD-biax-ell-Ey}.
When $\koc$ is small, the sole null of $Q_D(\theta,\phi)$ occurs close to $\theta=90^\circ$ 
in Figs. \ref{QD-biax-ell-Ex}(a) and \ref{QD-biax-ell-Ey}(b). This null can be attributed to 
the \magenta{electric-dipole} terms in Eqs.~(\ref{2}) and (\ref{2-H}). The contributions of the \magenta{magnetic-dipole} terms are
much smaller because $\mur$ is much closer to unity than $\epsr$ is; indeed, on interchanging the values
of $\epsr$ and $\mur$, we found the null to occur in Figs.~\ref{QD-biax-ell-Ex}(b) and \ref{QD-biax-ell-Ey}(a).
The contributions of the
higher-order multipole terms are vanishingly small because $\koc$ is sufficiently small. We have verified that the null
identified in Figs. \ref{QD-biax-ell-Ex}(a) and \ref{QD-biax-ell-Ey}(b) occurs exactly at 
$\theta=90^\circ$ when $\koc < 0.1$, just as for a  biaxially dielectric-magnetic sphere \cite{Jafri}.
As the electrical size of the scattering object increases, the formation of lobes in the scattering pattern is evident from 
the presence of multiple nulls in these figures, just like for isotropic-dielectric objects \cite{Bohren,Bowman,Asano,Barber}.
The foregoing remarks also apply to the scattering patterns of the chosen uniaxially dielectric-magnetic  spheroid  shown in Fig.~\ref{QD-uniax-sph-Ex}.

A  feature expected for the  biaxially dielectric-magnetic ellipsoid is that the scattering patterns for 
\begin{itemize}
\item
$\phi=0^\circ$ when $\hat{e}_{inc}=\hat{x}$ and
\item 
$\phi=90^\circ$ when $\hat{e}_{inc}=\hat{y}$
\end{itemize}
do not coincide.  This expectation, which emerges both from the   ellipsoidal shape of the object and
its constitutive anisotropy, is borne out in Figs. \ref{QD-biax-ell-Ex}(a) and \ref{QD-biax-ell-Ey}(b).
For the same reasons,
the scattering patterns for
$\phi=0^\circ$  in Fig.~\ref{QD-biax-ell-Ey}(a) do not coincide with
the scattering patterns for
$\phi=90^\circ$  in Fig. \ref{QD-biax-ell-Ex}(b). As both $a=b$ and  $\alpha_x=\alpha_y$ for the uniaxially dielectric-magnetic  spheroid, neither of the two
features is exhibited by the scattering patterns  in Fig.~\ref{QD-uniax-sph-Ex}.

\subsection{Total scattering and absorption efficiencies}
\begin{figure}
  \centering
     \begin{subfigure}[h]{0.4\textwidth}
\includegraphics[width=\linewidth]{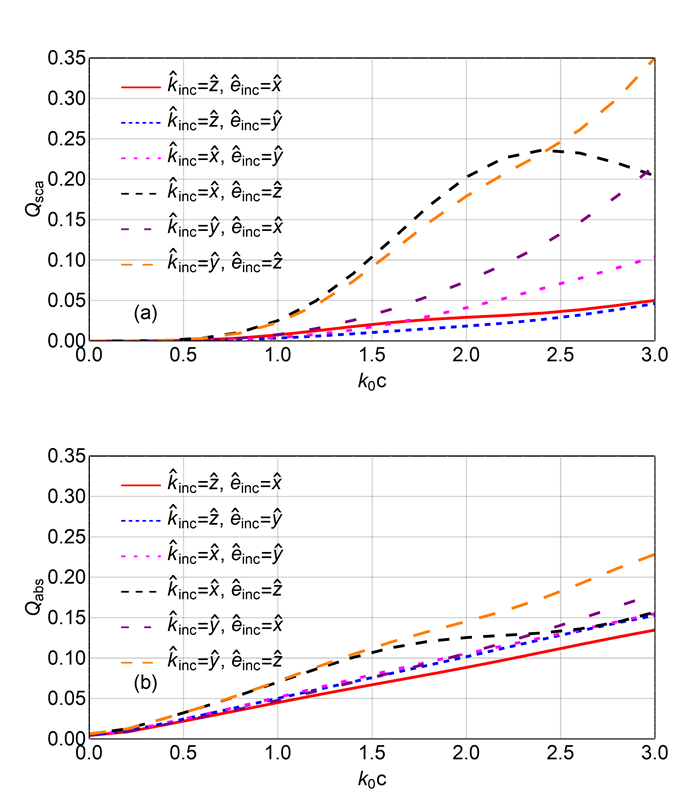}
 \end{subfigure}
\caption{(a) $Q_{sca}$ and (b) $Q_{abs}$  vs. $\koc$ for a biaxially dielectric-magnetic ellipsoid; $\epsr=2(1+0.1i)$, 
$\mur=1.05$, $\alpha_x=1.1$, $\alpha_y=1.2$, $a/c=1/2$, and $b/c=2/3$. 
}
\label{Qsca-Qabs-biax-ell}
\end{figure}

The total scattering efficiency $Q_{sca}$ and the absorption efficiency $Q_{abs}$ are plotted in 
Fig.~\ref{Qsca-Qabs-biax-ell}
as functions of $\koc$
for a biaxial dielectric-magnetic ellipsoid described by $\epsr=2(1+0.1i)$, 
$\mur=1.05$, $\alpha_x=1.1$, $\alpha_y=1.2$, $a/c=1/2$, and $b/c=2/3$. These results are shown for all six canonical configurations of the incident plane wave with respect to the semi-axes of the ellipsoid; i.e., 
$\hat{k}_{inc}\in\lec\hat{x},\hat{y},\hat{z}\ric$ and $\hat{e}_{inc}\in\lec\hat{x},\hat{y},\hat{z}\ric$ such that $\hat{k}_{inc}\perp\hat{e}_{inc}$.

Clearly, $Q_{sca}<Q_{abs}$ when $\hat{k}_{inc}=\hat{z}$, regardless of
the electrical size $\koc$ and  the polarization state of the incident plane wave.
Both $\hat{e}_{inc}$ and $\hat{h}_{inc}$ are then parallel to either $\hat{x}$ or $\hat{y}$, i.e.,
\begin{itemize}
\item
neither to the eigenvector of $\doubleunderline{\eps}_r$ (also, $\doubleunderline{\mu}_r$)
corresponding to its eigenvalue with the largest magnitude
\item
nor to the eigenvector of $\doubleunderline{U}$
corresponding to its largest eigenvalue.
\end{itemize}
Thus, $Q_{sca}<Q_{abs}$ for two canonical configurations of the incident plane wave in Fig.~\ref{Qsca-Qabs-biax-ell}. Calculations for $c/a=2$ and $b/a=2/3$ (results not shown here) indicate that $Q_{sca}<Q_{abs}$ holds regardless  of 
$\koc$ when $\hat{k}_{inc}=\hat{x}$. 
Thus, it would appear that  $Q_{sca}<Q_{abs}$ when $\hat{k}_{inc}$ is parallel to the eigenvector of $\doubleunderline{U}$
corresponding to its largest eigenvalue. However, calculations for 
$\epsr=4(1+0.1i)$, 
$\mur=1.1$, $a/c=1/2$, and $b/c=2/3$ (results not shown here) indicate that the inequality $Q_{sca}<Q_{abs}$ depends on $\koc$, even when $\hat{k}_{inc}=\hat{z}$; indeed, that inequality  holds for $\koc \leq 1.1$ when $\hat{e}_{inc}=\hat{x}$, and for $\koc \leq 1.4$ when $\hat{e}_{inc}=\hat{y}$.

The inequality $Q_{sca}<Q_{abs}$ does not hold in Fig.~\ref{Qsca-Qabs-biax-ell}
when either $\hat{e}_{inc}$ or $\hat{h}_{inc}$ is aligned
parallel to $\hat{z}$, i.e., the eigenvector corresponding to the  eigenvalues of
$\doubleunderline{\eps}_r$, $\doubleunderline{\mu}_r$, and $\doubleunderline{U}$
with the largest magnitude. Indeed, $Q_{sca}>Q_{abs}$ for smaller $\koc$ when $\hat{e}_{inc}=\hat{z}$
but for larger $\koc$ when $\hat{h}_{inc}=\hat{z}$.  Calculations for 
$\epsr=1.05$ and
$\mur=2(1+0.1i)$ (results not shown here) indicate that $Q_{sca}>Q_{abs}$ for larger $\koc$ when $\hat{e}_{inc}=\hat{z}$
but for smaller $\koc$ when $\hat{h}_{inc}=\hat{z}$.

\begin{figure}
 \centering
     \begin{subfigure}[h]{0.4\textwidth}
\includegraphics[width=\linewidth]{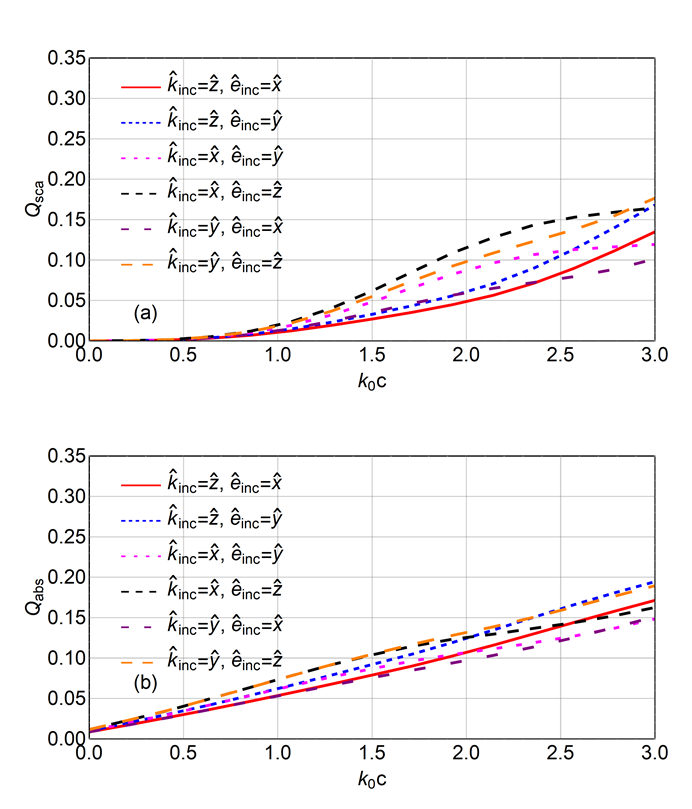}
 \end{subfigure}
\caption{Same as Fig.~\ref{Qsca-Qabs-biax-ell}, except that $\epsr=1.68(1+0.1i)$, 
$\mur=0.882$, and $\alpha_x=\alpha_y=1$.
}
\label{Qsca-Qabs-iso-ell}
\end{figure}

In order to understand the effect of shape alone, we repeated the calculations for Fig.~\ref{Qsca-Qabs-biax-ell}
but with $\epsr=1.68(1+0.1i)$, 
$\mur=0.882$, and $\alpha_x=\alpha_y=1$. Both $Q_{sca}$ and  $Q_{abs}$ are plotted in 
Fig.~\ref{Qsca-Qabs-iso-ell}
as functions of $\koc$
for this isotropic dielectric-magnetic ellipsoid whose relative permittivity  is the average of the three
eigenvalues of the relative permittivity dyadic and whose relative permeability is the average of the three
eigenvalues of the relative permeability dyadic  used for Fig.~\ref{Qsca-Qabs-biax-ell}.  When $\hat{k}_{inc}=\hat{z}$, $Q_{sca}$ for the isotropic dielectric-magnetic ellipsoid is greater than $Q_{sca}$ for the biaxially dielectric-magnetic ellipsoid, regardless of the electrical size $\koc$. This inequality does not hold for all $\koc$ when either $\hat{e}_{inc}=\hat{z}$ or $\hat{h}_{inc}=\hat{z}$; i.e., the eigenvector corresponding to the  eigenvalue of $\doubleunderline{U}$ with the largest magnitude. This inequality breaks down for smaller $\koc$ when $\hat{e}_{inc}=\hat{z}$
but for larger $\koc$ when $\hat{h}_{inc}=\hat{z}$. This breakdown can only be attributed to $\doubleunderline{U}\ne
\doubleunderline{I}$ because the material is isotropic.
Finally, $Q_{abs}$ for the isotropic dielectric-magnetic ellipsoid does not differ significantly from $Q_{abs}$ for the biaxially dielectric-magnetic ellipsoid. That is, the shape has a much more appreciable impact on $Q_{sca}$ than on $Q_{abs}$.

\subsection{Forward-scattering efficiency}

\begin{figure}
  \centering
     \begin{subfigure}[h]{0.4\textwidth}
\includegraphics[width=\linewidth]{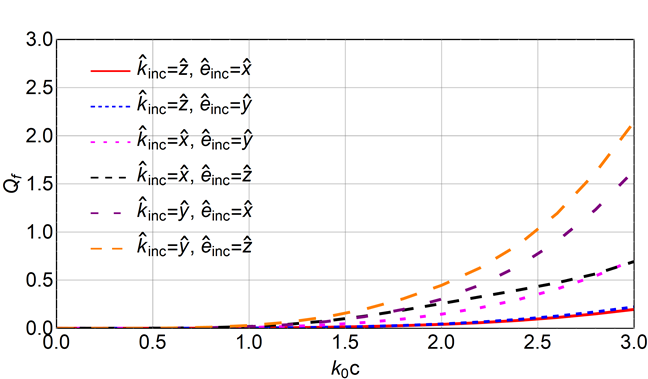}
 \end{subfigure}
\caption{$Q_{f}$  vs. $\koc$ for a biaxially dielectric-magnetic ellipsoid; $\epsr=2(1+0.1i)$, 
$\mur=1.05$, $\alpha_x=1.1$, $\alpha_y=1.2$, $a/c=1/2$, and $b/c=2/3$.  
  }
\label{Qf-biax-ell}
\end{figure}

For all six canonical configurations of the incident plane wave identified in Fig.~\ref{Qsca-Qabs-biax-ell}
for a biaxially dielectric-magnetic ellipsoid,
the forward-scattering efficiency $Q_f$ is almost a monotonically increasing function of $\koc \leq3$. This is clear
from Fig.~\ref{Qf-biax-ell} for   $\epsr=2(1+0.1i)$, 
$\mur=1.05$, $\alpha_x=1.1$, $\alpha_y=1.2$, $a/c=1/2$, and $b/c=2/3$. The same conclusion was drawn
for a uniaxially dielectric--magnetic spheroid  with $\alpha_x= \alpha_y=1.1$ and $a/c=b/c=1/2$ (results not shown here).

\subsection{Backscattering efficiency}

\begin{figure}
  \centering
     \begin{subfigure}[h]{0.4\textwidth}
\includegraphics[width=\linewidth]{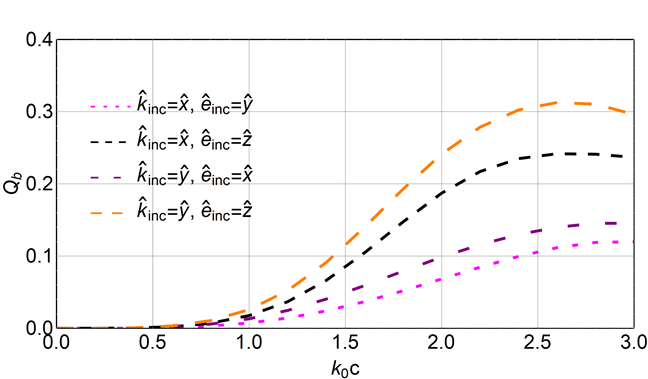}
 \end{subfigure}
\caption{$Q_{b}$  vs. $\koc$ for a biaxially dielectric-magnetic ellipsoid 
when $\hat{k}_{inc}\in\lec\hat{x},\hat{y}\ric$; $\epsr=2(1+0.1i)$, 
$\mur=1.05$, $\alpha_x=1.1$, $\alpha_y=1.2$, $a/c=1/2$, and $b/c=2/3$. 
  }
\label{Qb-biax-ell-kx-ky}
\end{figure}

When $\hat{k}_{inc}\in\lec\hat{x},\hat{y}\ric$,  the backscattering efficiency $Q_b$ is almost a monotonically increasing function of $\koc \leq3$, as shown in Fig.~\ref{Qb-biax-ell-kx-ky} for $\epsr=2(1+0.1i)$, 
$\mur=1.05$, $\alpha_x=1.1$, $\alpha_y=1.2$, $a/c=1/2$, and $b/c=2/3$. The same conclusion was drawn
for a uniaxially dielectric--magnetic spheroid  with $\alpha_x= \alpha_y=1.1$ and $a/c=b/c=1/2$ (results not shown here).

\begin{figure}
  \centering
     \begin{subfigure}[h]{0.4\textwidth}
\includegraphics[width=\linewidth]{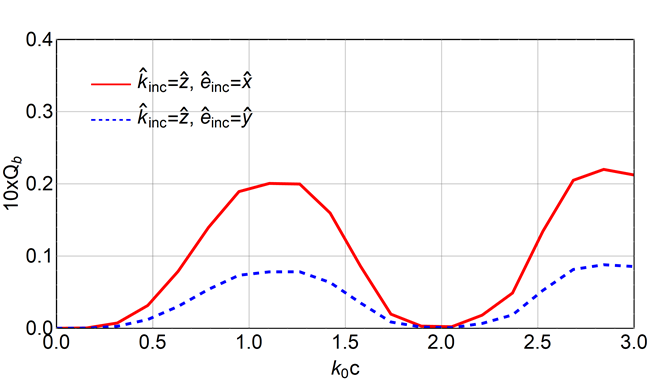}
 \end{subfigure}
\caption{$Q_{b}$  vs. $\koc$ for a biaxially dielectric-magnetic ellipsoid 
when $\hat{k}_{inc}=\hat{z}$; $\epsr=2(1+0.1i)$, 
$\mur=1.05$, $\alpha_x=1.1$, $\alpha_y=1.2$, $a/c=1/2$, and $b/c=2/3$. 
}
\label{Qb-biax-ell-Ex-Ey}
\end{figure}

In Fig.~\ref{Qb-biax-ell-Ex-Ey}, the variation of $Q_b$ with $\koc$ has an undulating character when 
$\hat{k}_{inc}=\hat{z}$. This is
in contrast to the monotonic increase for $\hat{k}_{inc}\in \{ \hat{x},\hat{y} \}$
in Fig.~\ref{Qb-biax-ell-kx-ky}.

\begin{figure}
  \centering
     \begin{subfigure}[h]{0.4\textwidth}
\includegraphics[width=\linewidth]{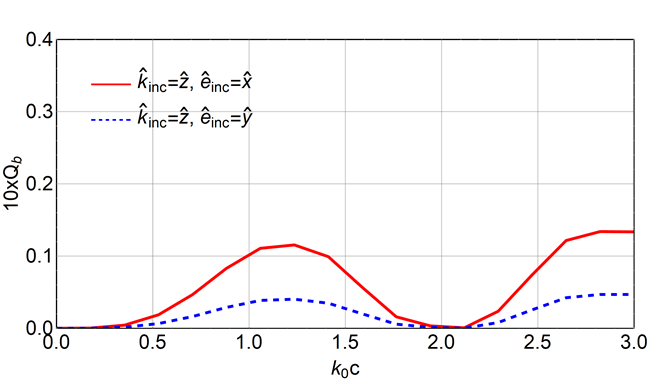}
 \end{subfigure}
\caption{Same as Fig.~\ref{Qb-biax-ell-Ex-Ey} except that $a/c=b/c=1/2$.
}
\label{Qb-biax-sph-Ex-Ey}
\end{figure}

The dependence of $Q_b$ in Fig.~\ref{Qb-biax-ell-Ex-Ey} on the
polarization state of the incident plane wave is strong. Indeed, $Q_b$ for $\hat{e}_{inc}=\hat{x}$ exceeds
$Q_b$ for $\hat{e}_{inc}=\hat{y}$. This must be due to both the shape ($a <b$) 
 and the constitutive anisotropy ($\alpha_x<\alpha_y$) of 
the scattering object.  When $\alpha_y<\alpha_x$ (results not shown here), $Q_b$ for $\hat{e}_{inc}=\hat{y}$ exceeds
$Q_b$ for $\hat{e}_{inc}=\hat{x}$. Furthermore, when $b<a$ (results not shown here), $Q_b$ for $\hat{e}_{inc}=\hat{x}$ exceeds
$Q_b$ for $\hat{e}_{inc}=\hat{y}$. Together, these data indicate that the constitutive anisotropy has a more
significant impact than shape on $Q_b$.

\begin{figure}
  \centering
     \begin{subfigure}[h]{0.4\textwidth}
\includegraphics[width=\linewidth]{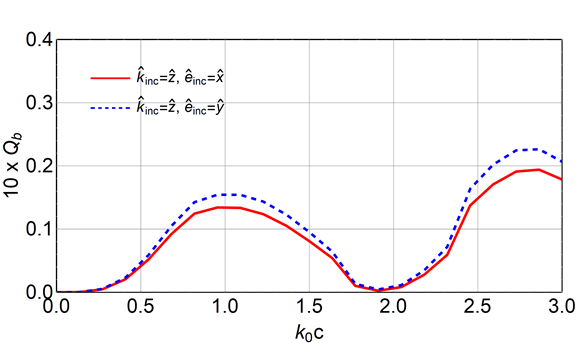}
 \end{subfigure}
\caption{Same as Fig.~\ref{Qb-biax-ell-Ex-Ey} except that $\alpha_x= \alpha_y=1.1$. 
}
\label{Qb-uniax-ell-Ex-Ey}
\end{figure}

The backscattering efficiency reduces in Fig.~\ref{Qb-biax-sph-Ex-Ey}
when both $a/c$ and $b/c$ are equal to $1/2$ but $\alpha_x<\alpha_y$, 
but
$Q_b$ for $\hat{e}_{inc}=\hat{x}$ still exceeds
$Q_b$ for $\hat{e}_{inc}=\hat{y}$.
The backscattering efficiency also reduces in Fig.~\ref{Qb-uniax-ell-Ex-Ey}
when both $\alpha_x$ and $\alpha_y$ are equal to $1.1$ but $a<b$, 
but now
$Q_b$ for $\hat{e}_{inc}=\hat{x}$ is somewhat lower than
$Q_b$ for $\hat{e}_{inc}=\hat{y}$. 
When $b<a$ (results not shown here), however, $Q_b$ for $\hat{e}_{inc}=\hat{y}$ becomes somewhat lower than
$Q_b$ for $\hat{e}_{inc}=\hat{x}$.  Finally, when both $\alpha_x=\alpha_y=1.1$ 
and $a/c=b/c=1/2$, the scattering object becomes a uniaxially dielectric-magnetic spheroid
and $Q_b$ does not depend on the polarization state of the incident plane wave.

\begin{figure}
  \centering
     \begin{subfigure}[h]{0.4\textwidth}
\includegraphics[width=\linewidth]{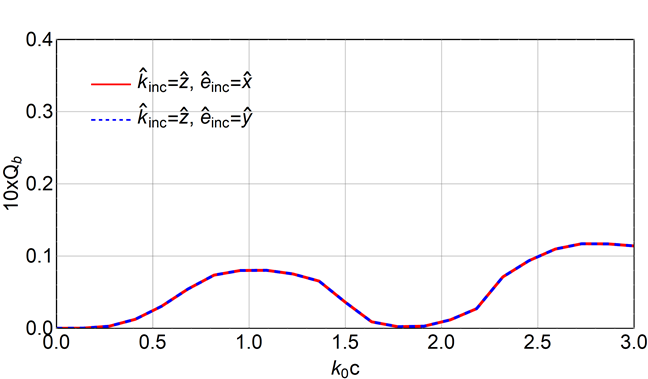}
 \end{subfigure}
\caption{Same as Fig.~\ref{Qb-biax-ell-Ex-Ey} except that $\alpha_x= \alpha_y=1.1$ and $a/c=b/c=1/2$.
}
\label{Qb-uniax-sph-Ex-Ey}
\end{figure}

One common observation in all the foregoing results for $\hat{k}_{inc}=\hat{z}$ is that the backscattering efficiency reduces to a very small value when ${\ko}c=2$. Zero backscattering efficiency has been previously reported
for isotropic dielectric-magnetic bodies of revolution when the incident plane wave  propagates parallel to the axis
of revolution and the scattering object is impedance matched to the free space surrounding it \cite{Wagner,Uslenghi3}.
However, for Figs.~\ref{Qb-biax-ell-Ex-Ey}--\ref{Qb-uniax-sph-Ex-Ey}, the scattering object is not necessarily a body of revolution, it is not isotropic, and it is not impedance matched to free space. Scattering objects exhibiting $Q_b\to 0$
but large $Q_f$ hold promise to enhance the detection and harvesting of incident electromagnetic energy \cite{Zhang}.

\section{Concluding Remarks}\label{s3}

We used the  extended boundary condition method  to study planewave scattering by a nonspherical object
composed of an orthorhombic dielectric-magnetic material whose relative permittivity dyadic is a scalar multiple of its relative permeability dyadic. Numerical results were obtained for scattering by ellipsoids with semi-axes aligned parallel to the eigenvectors
of the relative permittivity dyadic, hence allowing us to understand the relative impacts of constitutive anisotropy and nonsphericity.

Regardless of the direction of propagation and the polarization state of the incident plane wave,
the adequate number of terms
in the expansions of the scattered field phasors increase as the electrical size of the ellipsoid increases;
as a result, more lobes appear in the scattering patterns. In this respect, constitutive anisotropy cannot
be distinguished  from isotropy. However, constitutive anisotropy is inimical to symmetry in scattering
patterns.

The absorption efficiency can be either smaller or larger than the total scattering efficiency, depending on the
electrical size of the scattering object,  the ratio $\epsr/\mur$, and  the direction of propagation and the polarization state of the incident plane wave.
The shape of the scatterer has a more pronounced impact on  the total scattering efficiency than on 
the absorption efficiency.

Regardless of the configuration of the incident plane wave, the forward
scattering efficiency increases monotonically with the electrical size. The same characteristic
is displayed by the backscattering efficiency for some, but not all, planewave configurations.
For other configurations, the backscattering efficiency has an undulating behavior with increase in electrical size, and is highly affected by the shape and the constitutive anisotropy of the ellipsoid. The backscattering efficiency can be minuscule even when the forward-scattering efficiency is not, a desirable feature for harvesting incident
electromagnetic energy. Minuscule backscattering can also provide immunity from detection by monostatic
detection systems.

\vspace{0.5cm}
\noindent {\bf Acknowledgment.}  AL thanks the Charles Godfrey Binder Endowment at Penn State for ongoing support of his research
activities.

\section*{Appendix} The polarizability dyadic of an electrically small biaxial-dielectric ellipsoid
in vacuum is analytically known \cite{Lakh-BBPC}. Accordingly, 
\begin{itemize}
\item[(i)]
the polarizability dyadic of an electrically small
sphere of radius $R$ composed of a material with relative permittivity dyadic
$\eps_r^{x}\,\ux\ux+\eps_r^{y}\,\uy\uy+\eps_r^{z}\,\uz\uz$ and
\item[(ii)]
the polarizability dyadic of an electrically small
ellipsoid of semi-axes $a$, $b$, and $c$ and composed of a material with relative permittivity scalar
$\eps_r$ 
\end{itemize}
are identical, provided that
\begin{equation}
R^3=abc
\label{Rabc}
\end{equation}
and
\begin{equation}
\left.\begin{array}{l}
\eps_r^{x}= \displaystyle{
\frac{\eps_r\left(3L_x+2\right)+1-3L_x}
{\eps_r\left(3L_x-1\right)+4-3L_x}
}
\\[8pt]
\eps_r^{y}= \displaystyle{
\frac{\eps_r\left(3L_y+2\right)+1-3L_y}
{\eps_r\left(3L_y-1\right)+4-3L_y}
}
\\[8pt]
\eps_r^{z}= 
\displaystyle{
\frac{\eps_r\left(3L_z+2\right)+1-3L_z}
{\eps_r\left(3L_z-1\right)+4-3L_z}
}
\end{array}\right\}\,,
\label{eps-xyz}
\end{equation}
where \cite{Stoner,Osborn}
\begin{equation}
\left.\begin{array}{l}
L_x=\displaystyle{\frac{abc}{2}\int_0^\infty \frac{dq}
{(q+a^2)^{3/2} (q+b^2)^{1/2} (q+c^2)^{1/2}}}
\\[10pt]
L_y=\displaystyle{\frac{abc}{2}\int_0^\infty \frac{dq}
{(q+a^2)^{1/2} (q+b^2)^{3/2} (q+c^2)^{1/2}}}
\\[10pt]
L_z=\displaystyle{\frac{abc}{2}\int_0^\infty \frac{dq}
{(q+a^2)^{1/2} (q+b^2)^{1/2} (q+c^2)^{3/2}}}
\end{array}\right\}\,.
\label{L-xyz}
\end{equation}

Analogously  \cite{Lakh-BBPC},
\begin{itemize}
\item[(i)]
the magnetizability dyadic \cite{LMaeu,Krowne-Shen,Welter} of an electrically small
sphere of radius $R$ composed of a material with relative permeability dyadic
$\mu_r^{x}\,\ux\ux+\mu_r^{y}\,\uy\uy+\mu_r^{z}\,\uz\uz$ and
\item[(ii)]
the magnetizability dyadic of an electrically small
ellipsoid of semi-axes $a$, $b$, and $c$ and composed of a material with relative permeability scalar
$\mu_r$ 
\end{itemize}
are identical, provided that Eq.~(\ref{Rabc}) and conditions analogous to Eqs.~(\ref{eps-xyz})
and (\ref{L-xyz}) hold.

\end{document}